\documentclass{vgtc}                 

\vgtccategory{Research}        

\usepackage[table,svgnames]{xcolor}

\usepackage{times}                     
\usepackage{mathptmx}                  

\usepackage{booktabs}                  
\usepackage{array}
\usepackage{hyphenat}
\usepackage{float}

\title{A Comparative Study of Static, Scrollytelling, and Chatbot Visualization Onboarding Techniques for UX Designers}

\author{%
\texorpdfstring{%
\begin{tabular}{c}
\begin{tabular}{ccc}
Ester Chen & Aboli Shete & Aditya Anavekar \\
\texttt{esterchen@mail.rit.edu} & \texttt{as6151@rit.edu} & \texttt{aa3278@rit.edu}
\end{tabular}\\[0.8em]
\begin{tabular}{cc}
Roshan Peiris & Hidy Kong \\
\texttt{roshan.peiris@rit.edu} & \texttt{hidy.kong@rit.edu}
\end{tabular}\\[1.4em]
Rochester Institute of Technology\\
\end{tabular}%
}{Ester Chen, Aboli Shete, Aditya Anavekar, Roshan Peiris, Hidy Kong}%
}

\abstract{
User experience (UX) designers face barriers when creating data visualizations due to limited domain expertise in visualization or unfamiliarity with specialized tools. This highlights a clear need for effective methods to build visualization literacy. To address this, we evaluated three visualization onboarding techniques --- static, scrollytelling, and chatbot --- in an experimental study with 25 UX designers and students. We measured visualization comprehension and guideline adherence during a visualization creation task, followed by surveys and interviews to capture preferences and experiences. Compared to static onboarding, the pooled interactive condition (scrollytelling or chatbot) was associated with significantly higher guideline-adherence scores during visualization creation; both interactive techniques also received higher engagement ratings. Instruction clarity ratings were significantly higher when the two interactive conditions were pooled. Comprehension did not differ significantly across conditions. While participants generally preferred the interactive techniques, no significant differences emerged between scrollytelling and chatbot in performance or onboarding experience ratings. Drawing on the findings, we discuss three design dimensions of visualization onboarding (narrative structure, visual content layout, and navigational flexibility), their design implications, and potential opportunities for future research in this field.
}

\keywords{Visualization literacy, UX design, interactive learning, scrollytelling, visual onboarding, comparative study}
\begin{document}
\firstsection{Introduction}

\maketitle

As data becomes increasingly complex and multidimensional, traditional visualizations like bar and pie charts often fall short of capturing the full scope of the information \cite{keim2002information, few2009now}. To address this, advanced visualization techniques such as heatmaps, treemaps, and parallel set plots are required \cite{shneiderman1992tree, kosara2006parallel}. However, the effectiveness of these techniques is hindered by the prevalent lack of visualization literacy. Visualization literacy refers to ``the ability to confidently create and interpret visual representations of data'' \cite{boy2014principled}, which is essential in today’s data-driven world for effective communication and decision-making. However, non-experts including students and the general public, often struggle with complex visualizations \cite{borner2016investigating,boy2014principled,firat2022interactive}.

One such group of non-experts is user experience (UX) designers. A UX designer is a technical professional who focuses on creating a user-friendly experience for a product or service \cite{ixdf2024uxdesigner}. They often work with a variety of visualizations to create engaging dashboards that are tailored to the type of data and the dashboard's purpose. They also use data visualizations to communicate their research findings to stakeholders \cite{shashi2024journey}. UX designers responsible for dashboard design frequently face foundational challenges due to a lack of knowledge in this domain. Specifically, they face difficulties in identifying relevant data, choosing the appropriate visualization type, and effectively expressing the intended message \cite{bigelow2014reflections,Sumarsono2021,binhabib2023intersection,olohijere2023data}. Other issues include balancing aesthetic appeal and functionality, and ensuring that visualizations are accessible to all users, including those with disabilities. In addition, incorporating data visualization into UX design necessitates familiarity with a variety of tools and technologies \cite{Sumarsono2021,binhabib2023intersection,olohijere2023data}. Beyond individual skill gaps, the collaborative nature of dashboard creation introduces its own complexities. For instance, the process of handing off visualization designs to developers in multi-member teams frequently leads to design breakdowns, causing uncertainty and necessitating significant revisions during the development phase \cite{walny2019data}. When these handoff challenges are compounded by a UX designer's initial lack of visualization expertise, the risk of producing ineffective visualizations increases. Meanwhile, effective data visualization enables UX designers to clearly present complex data, improve user experience, engagement, and decision-making \cite{Sumarsono2021,binhabib2023intersection,olohijere2023data}. Given the existing challenges and anticipated benefits, there is a need to improve the visualization literacy of UX designers by helping them learn how to interpret and create effective data visualizations.

Visualization onboarding is an effective way to learn new visualization techniques as it is designed to help users comprehend and successfully create visual representations of data \cite{stoiber2019visualization, stoiber2021design}. There are various types of visualization onboarding techniques including textual tutorials, static tutorials (i.e., text and images), interactive tutorials, step-by-step guides, and video tutorials \cite{kwon2016comparative,stoiber2022comparative}.
Interactive tutorials have received positive feedback across a wide range of learning environments because of their ability to support core skills and self-paced learning \cite{Grant2006Developing, kwon2016comparative}. Despite their promise for supporting comprehension, little research has examined their effectiveness in developing the ability to \textit{create} visualizations. This study seeks to determine whether these techniques are equally beneficial for visualization onboarding for UX designers by comparing the impact of interactive tutorials and static tutorials on the designers' ability to both \textit{comprehend} and \textit{create} effective visualizations.

Interactive techniques like scrollytelling and chatbots have shown promise as learning tools through various studies \cite{stoiber2022comparative, perez2020rediscovering}. Scrollytelling is defined as a form of long-form storytelling that incorporates multimedia components including photos, movies, and interactive infographics with narrative text \cite{seyser2018scrollytelling}. This onboarding approach stands out for its ability to combine different types of content, resulting in an engaging and intuitive experience that compares favorably to traditional tutorials \cite{stoiber2022comparative, Morth2022ScrollyVis, Schneiders2020What}.
Chatbots provide another method of onboarding to increase user motivation and satisfaction by offering two-way interactive, personalized learning experiences. They can offer tailored e-learning support, allowing users to focus on relevant content while avoiding unnecessary material \cite{Klimova2023The, Riza2023Use}.
We chose scrollytelling and chatbot as the interactive techniques in our study given their effectiveness in other educational settings. The chatbot was rule-based with preset prompts and dialogues, rather than Artificial Intelligence (AI)-based, to strictly control experimental variables and ensure the informational content remained identical across all conditions. 
Throughout this paper, we refer to these two techniques collectively as interactive onboarding techniques.


Building on a preliminary survey of 139 participants to validate the study's necessity, we conducted a 1.5-hour comparative study with 25 UX design professionals and students. By contrasting static tutorials against scrollytelling and chatbot techniques, this work seeks to answer the following research questions:

\textbf{RQ1:} How do static, scrollytelling, and chatbot visualization onboarding techniques compare in their effectiveness in improving the ability of UX designers to \textit{comprehend} dense hierarchical data visualizations (specifically, heatmaps and treemaps)?


\textbf{RQ2:} How do static, scrollytelling, and chatbot visualization onboarding techniques influence UX designers' ability to correctly apply provided guidelines when they \textit{create} dense hierarchical data visualizations (specifically, heatmaps and treemaps)?

\textbf{RQ3:} How do interactive visualization onboarding techniques (scrollytelling, chatbot) influence UX designers’ perceived visualization onboarding experience in terms of confidence, engagement, information relevance, and instruction clarity?

The study results showed that the pooled interactive condition was associated with higher design guideline adherence during creation, and both interactive techniques received higher perceived engagement ratings than static onboarding. Perceived instruction clarity was higher when the two interactive conditions were pooled. We did not find a significant difference in comprehension results across onboarding techniques. Participants preferred interactive onboarding techniques over static onboarding because they perceived them as engaging and providing bite-sized information. Scrollytelling and chatbot techniques were comparable in all measured performance and rating metrics, with no significant differences in performance or perceived onboarding-experience ratings. Participants described distinct affordances for each technique, such as greater navigational flexibility in the chatbot and a smoother visual layout in scrollytelling.

Our study makes the following contributions:
\begin{itemize}
    \item Evaluating the effectiveness of three different types of onboarding techniques (static, scrollytelling, and chatbot) on UX designers' ability to \textit{create} visualizations according to design guidelines, and showing that the pooled interactive condition was associated with higher guideline-adherence scores.
    \item Comparing \textit{one-way and two-way onboarding techniques} in comprehension, visualization creation, and perceived onboarding experience, examining confidence ratings alongside participants' reports of navigational flexibility.
    \item Providing guidelines to create effective visualization onboarding based on three dimensions: narrative structure, information layout, and navigational flexibility.
\end{itemize}



\section{Related Work}
In this section, we cover prior work on visualization literacy, beginning with studies that demonstrate the need to increase literacy among the public and the broadening definition of visualization literacy. We then present works on evaluating and addressing visualization literacy and, finally, present research on visualization onboarding as a method to improve users' visualization comprehension and creation skills.

\subsection{Evaluating and Addressing Visualization Literacy}
Börner et al. state that in the era of information, the ability to interpret and create data visualizations is just as essential as traditional literacy skills such as reading and writing \cite{borner2016investigating}. Here, we distinguish between general visuospatial abilities and data-specific skills. The broader ability to understand and present information through a wide range of visuospatial representations (e.g., photographs, diagrams, maps, graphs) is known in cognitive psychology as \textit{graphicacy} \cite{tversky2001spatial}. In contrast, \textit{visualization literacy} more narrowly refers to "the ability to confidently create and interpret visual representations of data" \cite{boy2014principled}.\footnote{While the term ``visual literacy" is also used in the literature, we use ``visualization literacy" throughout the paper to avoid ambiguity, since visual literacy has been used interchangably with graphicacy as well.} Increasing visualization literacy improves communication and decision-making, and it also helps with professional development, educational advancement, and personal empowerment \cite{firat2022interactive, borner2019data}. Prior research demonstrated that teaching students to interpret and produce visual content improves their ability to communicate effectively and learn from visuals, emphasizing the importance of a visualization literacy curriculum as a stepping stone to data-specific skills \cite{tillmann2023competency}. However, research has also shown that the general public has lower levels of visualization literacy than is required to gain these benefits and fully utilize the power of data visualizations in everyday life \cite{firat2022interactive}.
Firat et al. conducted a literature review on visualization literacy \cite{firat2022interactive} and identified that various non-expert groups including MTurk workers \cite{kwon2016comparative, boy2014principled}, science museum visitors \cite{borner2016investigating, peppler2021cultivating}, elementary school students \cite{alper2017visualization,chevalier2018observations,bishop2019construct}, and higher education students \cite{firat2020treemap,wang2020cheat,rodrigues2021questions,krekhov2019integrating} faced difficulties when interpreting complex visual representations beyond basic charts. This gap calls for additional research on improving visualization literacy.

Although the importance of visualization literacy has long been recognized, its definition and scope have varied across the literature. Lee et al. \cite{lee2016vlat} defined it as “the ability and skill to \textit{read and interpret} visually represented data and extract information from data visualizations [emphasis added]". Solen's literature review on visualization literacy \cite{solen2022visualization} showed that most definitions similarly included keywords related to \textit{reading}, such as ``interpret,''  ``extract,''  ``comprehend,''  and ``understand''  \cite{borner2016investigating, chevalier2018observations, lee2016vlat, Lee2019VisualizationLiteracy}. In contrast, only a few definitions included \textit{writing}-oriented keywords such as ``create,'' ``design,'' and ``construct'' \cite{boy2014principled, huynh2021designing, peppler2021cultivating, alper2017visualization}.
This earlier focus on the \textit{interpretation} side is also reflected in numerous studies that proposed methods to measure and enhance visualization literacy \cite{ge2024toward, chevalier2018observations}. For example, the visualization literacy assessment test (VLAT) \cite{lee2016vlat} and Mini-VLAT \cite{Pandey_2023} are widely used to measure the visualization literacy of non-experts. However, these instruments focus solely on comprehension and overlook skills related to visualization creation. Highlighting this gap, Ge et al. called for broader assessments that include higher-level tasks (e.g., visualization construction) in measuring visualization literacy \cite{ge2024toward}.

A similar focus on interpretation can be found in studies on addressing low visualization literacy \cite{firat2022interactive, sarfraz2024, huynh2021designing, ashley2019improving}. For instance, Sarfraz et al. \cite{sarfraz2024} introduce \textit{Vizard}, an LLM-powered dashboard companion that explains existing visualizations in a user’s preferred language and complexity level, enabling tailored guidance for interpreting charts and extracting domain-relevant insights. In another study, reflective feedback mechanisms and interactive tools that guide users through the steps of visual reasoning processes have been shown to deepen understanding \cite{firat2022interactive}. Other works have explored innovative pedagogical approaches, such as gamification and narrative-focused role-playing games, and found them to be effective in helping learners, especially younger audiences, engage more deeply with and understand complex visual data \cite{huynh2021designing, ashley2019improving}. Another emerging strategy for improving visualization literacy is visualization onboarding, which refers to the process of teaching users how to read, interpret, and extract information from visual data representations, such as incorporating tutorials directly into visualization tools to support users with low visualization literacy \cite{firat2022interactive}.
Prior work has shown various types of visualization onboarding to be effective in enhancing visualization literacy in terms of comprehension \cite{stoiber2019visualization, stoiber2019visualization, kwon2016comparative}. For instance, Stoiber et al. \cite{stoiber2022abstract} compared concrete and abstract onboarding materials for teaching treemap,
finding that concrete examples can facilitate comprehension and insight generation more effectively than abstract descriptions alone \cite{stoiber2022abstract}. We review visualization onboarding techniques in more detail in the next subsection.

More recently, researchers have begun to recognize the importance of the \textit{creation} aspect of visualization literacy. Börner et al. introduced a framework for Data Visualization Literacy, which encompasses the ability to read, interpret, and \textit{construct} data visualizations effectively \cite{borner2019data}. Their assessment includes creation-based exercises, making a shift toward more comprehensive evaluation of visualization literacy \cite{borner2019data}. Similarly, Ge et al. presented an instrument that measures people's visual encoding ability in visualization construction, expanding the scope of visualization literacy \cite{ge2025avec}.
Researchers have also begun to explore techniques for improving visualization literacy with a focus on the creation of data visualizations. Structured education and training programs were designed to teach visualization design principles and skills \cite{stoiber2022abstract}, as well as hands-on practical exercises using real-world datasets where manually creating visualizations helps learners internalize the logic of visual encoding \cite{mailloux2008}. Creative and experiential approaches that use physical aids, such as 3D printed models \cite{lebow2018improving} and origami-based modules \cite{marji2023}, have also been shown to strengthen spatial reasoning and design abilities of learners.

More recently, ChatGPT-4 has been employed to teach data visualization design, overcoming traditional barriers related to instructor design expertise and software knowledge by using everyday language as an interface for visualization creation \cite{lear2023chatgpt4}.
Yet, existing research on improving visualization creation skills remains limited, as most methods were studied in educational settings, especially in K-12 environments, and often lack formal evaluation of their effectiveness.
Our work contributes to this line of work by examining the impact of onboarding techniques on visualization \textit{creation} in addition to comprehension, and broadening the focus beyond educational settings to professional contexts with UX designers.



\subsection{Visualization Onboarding Techniques}
While onboarding techniques have gained attention as a way to support users’ understanding of new visualizations \cite{stoiber2019visualization,firat2022interactive},
onboarding practices in the visualization field lack standardization, especially in the type of onboarding material (e.g., static text, video) that should be used during the process \cite{stoiber2019visualization}. To this end, researchers have begun to evaluate the effectiveness of various onboarding techniques in improving visualization comprehension.

Wang et al. created static, structured cheat sheets with visual guides and design guidelines to simplify and explain complex visualization techniques, improving accessibility and comprehension \cite{wang2020cheat}. Moving beyond static guides, more recent studies have introduced interactive onboarding techniques designed to support active learning and on-demand assistance. For instance, one study focused on helping analysts understand network patterns \cite{InteractivePattern}, where visuals were mapped to network patterns and the pattern explainer provided the corresponding visual and textual description. The users could select an area of the visualization and learn about it on demand. Results showed that interactive explanations not only improved participants’ ability to learn i) unfamiliar visualizations, ii) patterns in network science, and iii) network terminology, but also enhanced their comprehension, recall of concepts, and confidence in applying these visual analysis techniques.

Other studies have compared the effectiveness of different onboarding formats. Kwon and Lee compared four onboarding techniques --- baseline(textual), static, video, and interactive for individuals with no prior visualization experience \cite{kwon2016comparative}. Participants found video and interactive tutorials more engaging, easier to follow, and more enjoyable than static ones, with interactive tutorials being the most effective for learning and user engagement \cite{kwon2016comparative}. Similarly, Stoiber et al. conducted a study comparing three onboarding techniques for comprehension of complex data visualizations: step-by-step guides, video tutorials, and scrollytelling \cite{stoiber2022comparative}.
While there was no significant difference in answer correctness across conditions, participants had the quickest response times after watching the scrollytelling tutorial compared to when they viewed step-by-step instructions, the video tutorial, or did not have onboarding \cite{stoiber2019visualization}. In a follow-up study, Stoiber et al. compared two onboarding techniques for a visual analytics tool, in-situ scrollytelling onboarding that was built directly into the tool and external textual onboarding, and they found that participants preferred in-situ scrollytelling over external tutorials \cite{stoiber2022comparative}. Continuing this line of work, our study aims to compare the effectiveness of three onboarding techniques on how they support visualization literacy. We chose scrollytelling as one of the onboarding techniques based on prior findings highlighting its effectiveness as a visualization onboarding technique.

While prior work suggests that interactive onboarding techniques can be more engaging and effective than static or video-based techniques, it has focused mainly on one-way interactive onboarding where users engage with the content in a controlled, predetermined manner and follow pre-designed interactions. One possible approach to increase user control and engagement is through a two-way interactive technique, where users actively participate and influence the onboarding process, rather than merely following pre-designed interactions. A method for implementing a two-way interaction is through chatbots. Chatbots are computer programs designed to mimic human interaction, and have been shown to facilitate learning by providing real-time assistance, personalized lessons, increased engagement, self-paced options, and feedback \cite{wollny2021we}. An in-depth review by Perez et al. investigates the various types of educational chatbots and their impact on student learning and suggests that they could be used as a dynamic onboarding tool \cite{perez2020rediscovering}. Kong et al. have also suggested the use of a chatbot for visualization onboarding due to its interactive nature \cite{kong2022effects}. Given this potential of chatbots to allow users to actively engage in the onboarding process and enhance their perceived learning experience, we chose chatbots as the two-way interactive onboarding technique for our study.

In all, prior work suggests that interactive tutorials can support visualization comprehension and perceived engagement compared to static or textual methods. Our study advances this research domain by comparing the effectiveness of static, one-way, and two-way interactive onboarding techniques for both comprehension and creation of visualizations as well as perceived user experience.


\section{Method}
To answer our research questions, we conducted a preliminary survey with 139 respondents to learn about their challenges and interests in data visualization and ran a comparative study with 25 UX design professionals and students on three visualization onboarding techniques: static, chatbot, and scrollytelling. In this section, we describe the preliminary survey and the study procedure for the comparative study in more detail.


\subsection{Preliminary Survey}
Prior to designing the comparative study, we administered a preliminary survey to UX designers to identify their pain points and needs when designing data visualizations. The survey's primary goal was to learn about the designers' specific areas of interest in data visualization, their practical experience designing dashboard visualizations, and their perceptions of the importance of different types of visualizations. The preliminary survey was conducted using Qualtrics and distributed through various social media platforms, including LinkedIn, Discord, and Instagram, as well as through UX design courses offered by a university. The survey received 139 responses from UX designers with varying levels of experience.
When asked which visualizations designers need to learn for their work, 74\% of respondents expressed a particular interest in treemaps and heatmaps. Treemaps and heatmaps are widely used but often challenging to interpret due to their complexity \cite{Molina2023Effect, firat2020treemap}.  As a result, these visualization types were chosen as the focus of the onboarding tutorials in our study.
Participants self-reported their visualization familiarity and experience as part of the survey.
The survey results provided a valuable framework for understanding the current landscape of UX designers in data visualization by highlighting challenges such as the current struggles in creating complex visual dashboards using tools like Tableau, difficulties in collaborating with stakeholders to understand data needs, and designing user-friendly, insightful visualizations tailored to various industries. The results also validated the need for visualization onboarding tutorials which aimed to help designers overcome these challenges during a visualization creation task.

\subsection{Design of Visualization Onboarding Tutorials}
The study had three conditions based on the onboarding technique: (1) static, (2) chatbot, and (3) scrollytelling. For each of the onboarding conditions, we created two web-based tutorials, one about heatmaps and one about treemaps, resulting in six onboarding tutorials. Instructional content, datasets, and core visual examples remained consistent across conditions, while presentation and interaction style varied by onboarding technique. Each tutorial used the same three datasets: one for explanations, one for the quiz, and one for the creation tasks, ensuring consistency across all tutorials. The onboarding text was also the same across the three tutorials and was authored by the researchers, drawing on guidelines and examples from existing onboarding tutorials and other relevant references. The onboarding content can be found in the Supplementary Materials.

\subsubsection{Static Tutorial}
The static tutorial served as a baseline for evaluating the effectiveness of interactive onboarding tutorials. It was formatted as a plain text document, with accompanying images placed beneath their text descriptions (See Appendix \autoref{fig:static_left} to \autoref{fig:static_right}). The images in the static tutorial were screenshots from the scrollytelling version.

\subsubsection{Scrollytelling}
The Scrollytelling tutorial presented visualization design guidelines in a narrative flow. The implementation of the scrollytelling prototype closely followed the design and techniques used in interactive visualization articles on news websites such as New York Times articles (e.g., \cite{aisch2014worldcup}). It used a step-by-step format in which users scrolled through the tutorial, with each step introducing and explaining one guideline at a time. This method allowed users to gradually understand and see how each design principle was applied within the context of a coherent story. The prototype displayed explanatory text on the left side of the screen and corresponding visualization on the right (See Appendix Figure \ref{fig:Stop}). The visualization had an animated effect that changed as the user scrolled up and down. For example, in the treemap tutorial, the outer rectangle is initially divided into categories (See Appendix Figure \ref{fig:Sbottom_left}), then as the user scrolls down, it gets further divided into subcategories (See Appendix Figures \ref{fig:Sbottom_mid} and \ref{fig:Sbottom_right}), creating an animated transitional effect. To complete the tutorial, users scrolled through the tutorial, receiving information on how to create the visualization by engaging in a one-way interaction. The scrollytelling tutorial prototype was adapted from the implementation by Driessen \cite{driessen_scrollytelling_demo}, built with HTML, CSS, and JavaScript. The data visualizations were created with the D3.js library.

\subsubsection{Chatbot}
\begin{figure}[!ht]
    \centering
    \includegraphics[width=0.75\linewidth]{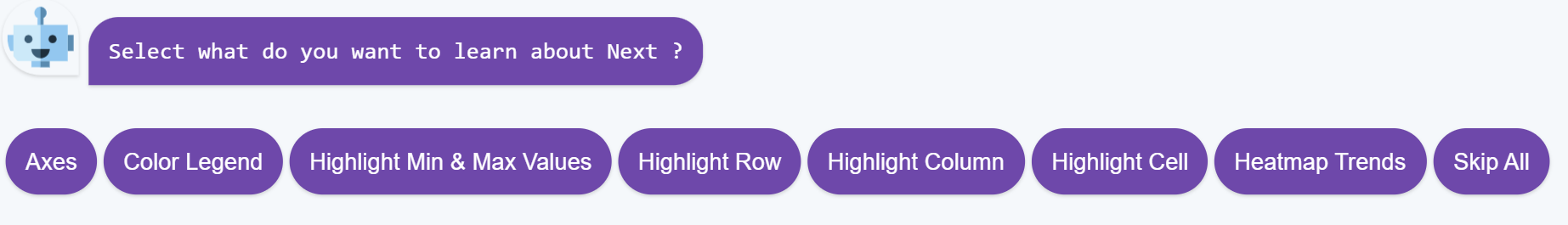}
    \caption{Preset prompts that users can choose from to select tutorial sections.}
    \label{fig:chatbotoptions}
\end{figure}

\begin{figure}[!ht]
    \centering
    \includegraphics[width=0.85\linewidth]{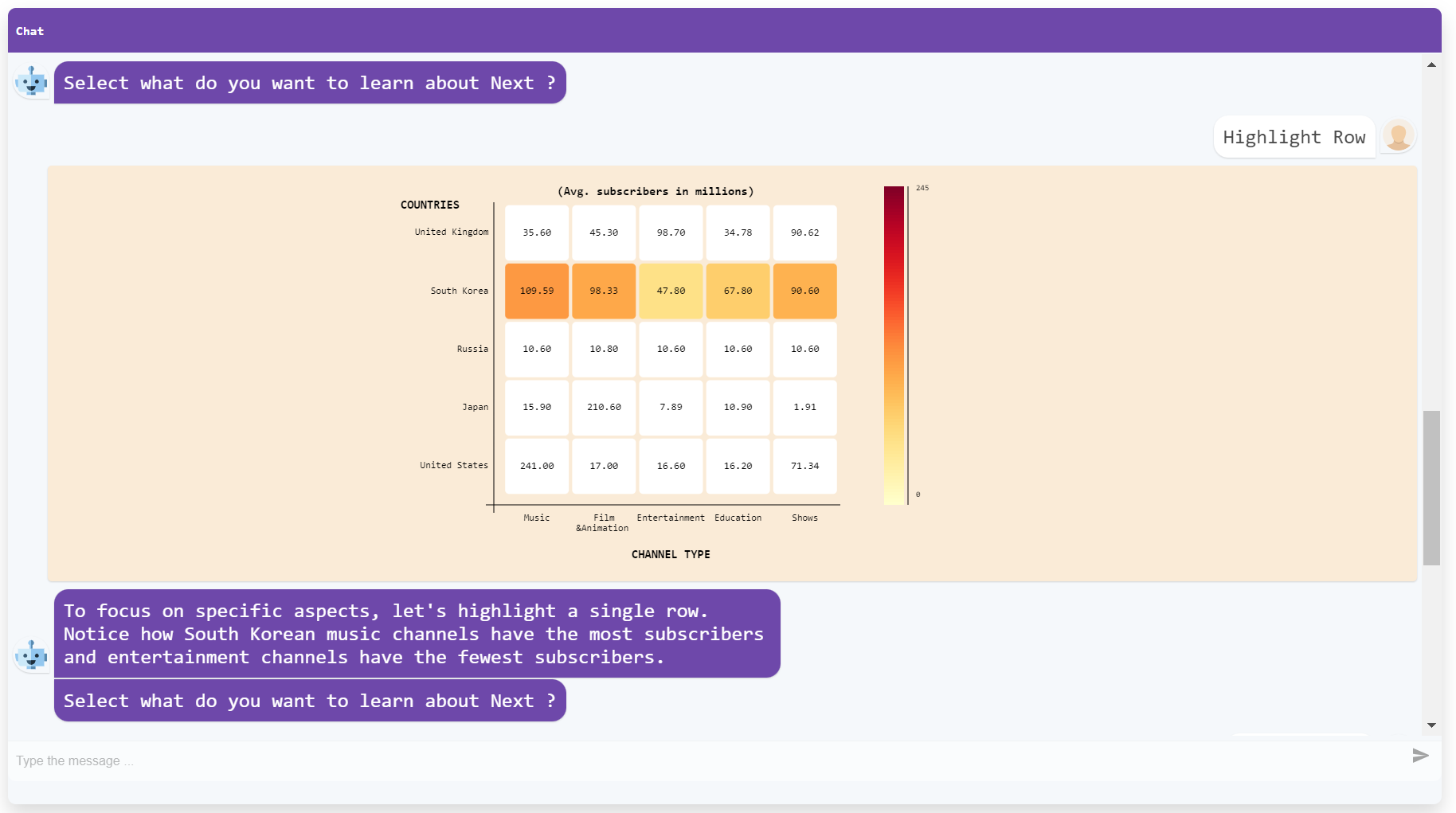}
    \caption{A step displayed dynamically in the chatbot tutorial, according to the user's selected option, including the relevant text and image.}
    \label{fig:chatbot}
\end{figure}

To tightly control the pedagogical narrative and ensure consistent delivery of the design guidelines, the chatbot in this study was implemented as a scripted, rule-based dialog system rather than a generative Large Language Model (LLM) or open-ended conversational AI. This prototype included preset prompts that enabled users to select which topics they wanted to explore next and skip steps they found irrelevant (See Figure \ref{fig:chatbotoptions}). The chatbot did not support free-form queries, but users could choose from various guideline options to receive detailed explanations on specific points (See Figure \ref{fig:chatbot}) or opt to skip directly to the quiz section if they preferred. After each step, users had the option to revisit any previous section if they wished to review or repeat content. This design supported user-controlled navigation through preset topic choices. For example, in the treemap tutorial, the learner could skip the section on labels if they are already familiar with it and instead begin with the section on coloring categories. This design enabled two-way interaction between the user and the chatbot.
The chatbot tutorial prototype was built with Node.js and Express on the backend. React-DOM, HTML, CSS, JavaScript, and a React-based chatbot library `react-simple-chatbot' were used on the frontend.

\subsection{Participants}
For the experimental study, twenty-five participants were drawn from a diverse subset of over 100 preliminary survey respondents.
The eligibility criteria for participation were prior experience in UX design and designing data visualizations for dashboards. All participants were user experience design professionals or students with varying levels of experience in both UX design and data visualization. We included UX designers with varying levels of experience, from novice to expert, to investigate how onboarding techniques influence designers based on their familiarity with UX design and data visualization principles. The participants were between 21 and 32 years old (\textit{M} = 26.5, \textit{SD} = 2.36) and included 11 males (44\%) and 14 females (56\%). Out of the 25 participants, 18 were MS students with a Bachelor's degree who were currently enrolled in a UX design related program, 2 were professionals with a Bachelor's degree, and 5 were professionals with a Master's degree. The study lasted for 1.5 hours and the participants were compensated with a \$20 gift card for their participation.

\begin{figure}[!htbp]
    \centering
    \includegraphics[width=\linewidth]{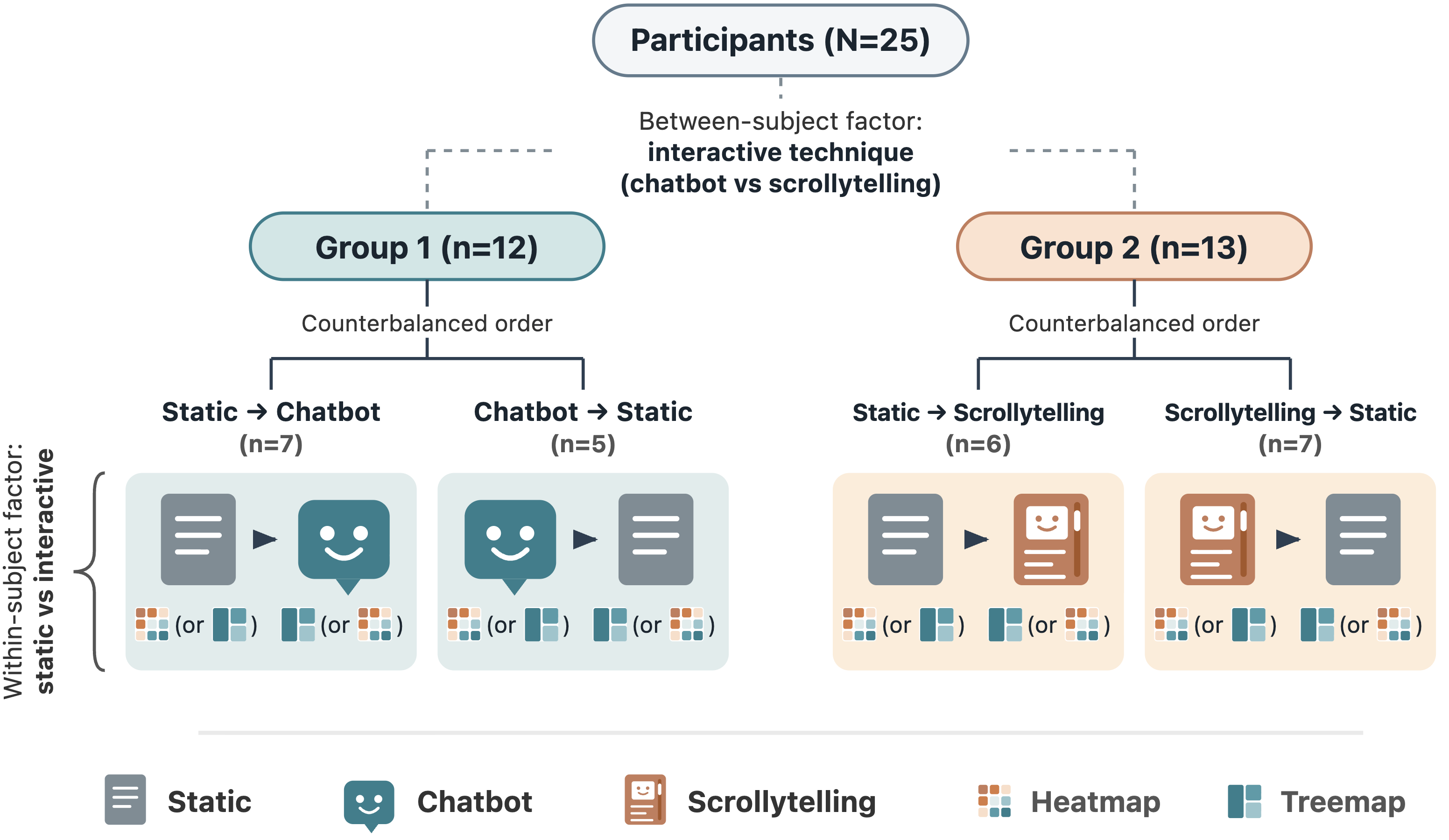}
    \caption{\textbf{Study Design Setup.} The study followed a mixed experimental design where \textit{between-subjects} factor was the interactive modality (Group 1: Chatbot; Group 2: Scrollytelling) and the \textit{within-subjects} factor was the onboarding modality (Static baseline vs. Interactive). To mitigate carry-over effects, the order of visualization types and onboarding methods were counterbalanced.}
    \label{study-design-flow}
\end{figure}

\subsection{Experimental Study}
Our experimental study employed a mixed experimental design as shown in \autoref{study-design-flow} to balance methodological rigor with practical feasibility. To compare the baseline (i.e., static) against interactive onboarding techniques, we utilized a within-subjects design; all participants experienced the static condition and one of the interactive conditions. We used a between-subjects design for the two interactive techniques, where roughly half of the participants used chatbot as the interactive technique and the other half used scrollytelling. Including all three conditions in a purely within-subjects design would have substantially increased the study duration, potentially leading to participant fatigue and negatively impacting performance.
Participants were divided into two groups (Group 1: chatbot, Group 2: scrollytelling), where each group had a diverse range of self-reported visualization literacy levels. Prior to the study, participants reported their familiarity with heatmap and treemap, and rated their knowledge of data visualization on a scale of 1 to 5 with a justification for their scores (score distributions overall and across experimental conditions are in the Supplementary Materials).

To mitigate learning effects, we counterbalanced the order of data visualization type (heatmap or treemap) across the two within-subjects conditions. For example, participants who started with the heatmap onboarding were assigned the treemap for their second onboarding tutorial, and vice versa. Furthermore, the presentation order of the onboarding conditions (e.g., Static first vs. Interactive first) was counterbalanced. The exact conditions assigned to each participant are provided in the Supplementary Materials. Each tutorial consisted of a brief explanation of the visualization, eight design guidelines for creating the visualization type, and a comprehension quiz (detailed further in Section 3.5).

After each tutorial, participants were asked to create a visualization in Miro\footnote{Miro is an online collaborative whiteboard platform: \url{https://miro.com}}. To evaluate their ability to recall and apply the guidelines covered in the tutorial without causing undue fatigue from building a visualization from scratch in Miro, participants were provided a Miro board pre-populated with a basic toolkit, including a dataset table, placeholder rectangles, and a color palette with diverse hues and color values (included in the Supplementary Materials). This setup mirrored a mid-fidelity UX wireframe task, allowing us to assess conceptual understanding and application of guidelines (e.g., using color saturation for numeric data), rather than technical design skills, such as a HEX code selection. We explicitly instructed participants that exact implementation details, such as exact pixel proportions for treemaps, were not the focus of the task. Instead, they were asked to focus on demonstrating conceptual knowledge and correct relative relationships (e.g., ensuring a larger data value was represented by a visually larger rectangle, or a higher value by a distinctly darker color shade). Participants were also asked to verbalize design intentions that were difficult to execute manually in Miro interface (e.g., many participants verbalized ``I would apply this for all other rows" after formatting a single row in heatmap).

After the task, they completed a post-onboarding survey to rate their overall experience with the tutorial. The process was repeated for the second onboarding tutorial. After the onboarding phase, we interviewed participants about their overall experience, particularly useful aspects, and the use of instructions during visualization creation. Participants were also asked about any challenges they encountered, their reasoning for the confidence rating in the post-onboarding survey, and potential tutorial improvements. Furthermore, feedback was requested on how the interactive nature of the scrollytelling and chatbot tutorials affected their learning process, as well as comparative thoughts on static and interactive techniques after completing the tutorials and visualization tasks.
Before starting the main study, two pilot studies were conducted to test the conditions for both groups. These pilots helped fine-tune the onboarding prototypes, and determine the time required to complete the tasks. The study procedure was approved by the Institutional Review Board, and informed consent was obtained from all participants.


\subsection{Evaluation}
The following measures were used to assess visualization comprehension, creation, and onboarding experience ratings.

\subsubsection{Comprehension Quiz}
Visualization comprehension was measured through a comprehension quiz administered at the end of each tutorial. To prevent task-learning effects, the quizzes used a different dataset from the one covered in the tutorial. The quiz was based on Kwon and Lee's Visualization Literacy Assessment Test \cite{lee2016vlat}. It consisted of five questions designed to assess participants' ability to recognize extremes, make comparisons, and retrieve values. For example, a value retrieval question asked, ``\textit{In which year did Australia have the smallest population compared to other years?}'' while a comparison question asked, ``\textit{Which Continent had a larger population than Asia in 2010?}'' Participants received a score out of five for each quiz.

\subsubsection{Visualization Guidelines}
To assess how effectively participants applied the concepts from their respective onboarding conditions during the creation task, we embedded eight essential design guidelines in each onboarding tutorial. These guidelines were based on existing guidelines for heatmaps and treemaps such as incorporating legends, proper axis labels, color scales based on data type, proportional rectangle sizing, and clear labeling and coloring of categories\cite{few2006information,ware2012information,kirk2016data,heer2010tour,tableauHeatmaps,few2012show,shneiderman1992tree}. To minimize learning effects, six of the eight guidelines for the heatmap and treemap tutorials were unique to the visualization type, while the other two were variants of the same design principle. The full list of guidelines for heatmaps and treemaps used in the tutorials can be found in our Supplementary Materials\footnote{Supplementary Materials: \url{https://anonymous.4open.science/r/vis-onboarding-1061}}.

Visualizations produced during the creation task were evaluated against these guidelines, with scores given based on the number of guidelines followed.
To evaluate guideline adherence, two researchers independently scored each participant-generated visualization against a predefined rubric based on the eight design guidelines. For example, we verified that higher numerical values in heatmaps were consistently mapped to darker colors than lower values. For treemaps, we verified that the hierarchical grouping was accurate and that the ordinal relationships of the rectangle sizes matched the data (i.e., larger values were assigned visibly larger areas).
After initial independent coding, the researchers held a consensus meeting to resolve ambiguities and achieved full agreement on all grading points.
To provide concrete examples of how the guidelines were applied or violated in practice, we provide a representative sample of participant-generated visualizations in the Supplementary Materials.

\subsubsection{Post-Onboarding Survey}
The post-onboarding survey included five Likert scale-based questions designed to elicit feedback from participants regarding their onboarding experience. It collected ratings of confidence in designing a visualization following the tutorial, engagement, instruction clarity, information relevance to participants' learning needs, and overall satisfaction.

\subsection{Data Analysis}
To analyze the quantitative data, we used paired $t$-tests for within-subject comparisons (static vs. interactive), examining comprehension quiz results and guideline adherence scores. Independent $t$-tests were used to compare between-subjects (scrollytelling vs. chatbot) using the same quantitative metrics. For Likert scale data, the Wilcoxon Signed-Rank test was used for within-subject comparisons (static vs. interactive), and the Mann-Whitney $U$ test for between-subject comparisons (scrollytelling vs. chatbot). These statistical methods allowed us to assess condition differences in participant performance and feedback across tutorial types. We also conducted linear regression analyses to examine whether participants’ prior visualization knowledge had an influence on participant scores (i.e., quiz and guideline adherence), beyond the effects of the onboarding type. We first ran an additive model (\texttt{score $\sim$ vis\_knowledge + type}) to assess independent contributions of each predictor, and then an interaction model (\texttt{score $\sim$ vis\_knowledge * type}) to explore whether the relationship between visualization knowledge and performance varied across onboarding types.

We analyzed the interview responses using thematic analysis \cite{braun2012thematic}, following a structured procedure. Initially, we transcribed the interview recordings to create detailed transcripts. Next, we developed initial themes from these transcripts. Using Miro, we reviewed and refined these themes collectively, identifying patterns and grouping related themes together. This collaborative process allowed us to ensure a comprehensive and refined understanding of the data, highlighting key insights and trends.

\section{Results}
\subsection{Quantitative Results}
First, we compare the visualization comprehension (i.e., quiz), creation (i.e., guideline adherence), and experience scores between two categories of visualization onboarding: (1) Static and (2) Interactive techniques. Subsequently, we present a comparison between two types of interactive onboarding techniques --- scrollytelling and chatbot.

\subsubsection{Interactive Onboarding is Associated with Higher Engagement and Guideline Adherence}
All participants experienced static onboarding and one interactive onboarding technique. Although higher quiz scores were observed in the pooled interactive condition (\(M{=}4.76,\,SD{=}0.43\)) compared to the static condition (\(M{=}4.48,\,SD{=}0.87\)), the difference was not statistically significant ($p = .08$) (See Appendix Figure \ref{fig:quantResults}a). In contrast, guideline adherence scores were significantly higher ($p = .04$, $d = 0.55$) in the pooled interactive condition (\(M{=}6.20,\,SD{=}1.63\)) than in the static condition (\(M{=}5.32,\,SD{=} 1.57\)) (Appendix Figure \ref{fig:quantResults}b), representing a medium effect size. Lower scores in static conditions were largely driven by structural and guideline errors; for example, a typical low-scoring submission frequently exhibited mistakes such as missing legends and value labels, which can make it difficult to infer values from the visualization (see supplementary materials for Miro board examples).
In addition to these comparisons, we examined the effect of prior visualization knowledge on quiz scores and guideline adherence scores. The results indicated that prior knowledge did not significantly account for differences in performance across participants.

\begin{figure}[!ht]
    \centering
    \includegraphics[width=\linewidth]{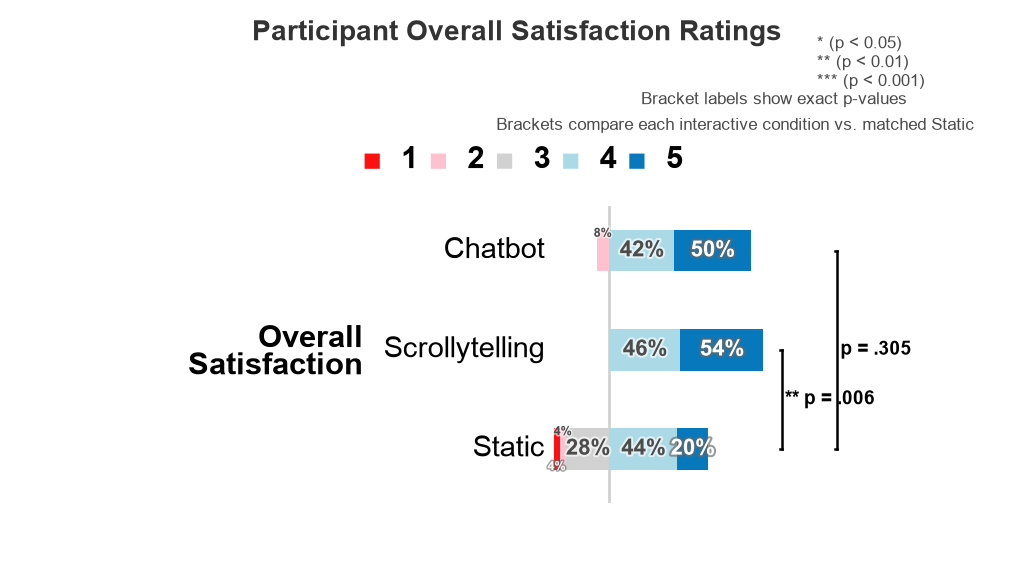} 
    \caption{Participants' overall satisfaction with the onboarding techniques, measured on a 5-point Likert scale (1 = Extremely Dissatisfied, 5 = Extremely Satisfied). Percent labels are rounded within each row to sum to 100\%. Consistent with the main results, satisfaction was significantly higher for scrollytelling compared to static, while chatbot did not significantly differ from static.}
    \label{fig:satis_likertResults}
\end{figure}

Likert scale survey responses revealed statistically significantly higher engagement ratings in the interactive conditions compared to the static condition. Instruction clarity ratings were significantly higher in the pooled interactive condition and for scrollytelling, but not for chatbot. Confidence ratings were numerically higher in the pooled interactive condition but did not reach the conventional significance threshold, and satisfaction ratings were significantly higher for scrollytelling than for static.
Engagement ratings were significantly higher for both the scrollytelling ($M = 4.23, SD = 0.60$; $p = .006$) and chatbot condition ($M = 4.42, SD = 0.67$; $p = .012$) compared to their matched static conditions (overall static mean for context: \(M{=}3.00,\,SD{=}1.32\)). Similarly, satisfaction was significantly higher for scrollytelling than for its matched static condition ($p < .01$), while chatbot did not significantly differ from its matched static condition ($p = .305$; Appendix Figure \ref{fig:satis_likertResults}).

\begin{figure}[!ht]
    \centering
    \includegraphics[width=\linewidth]{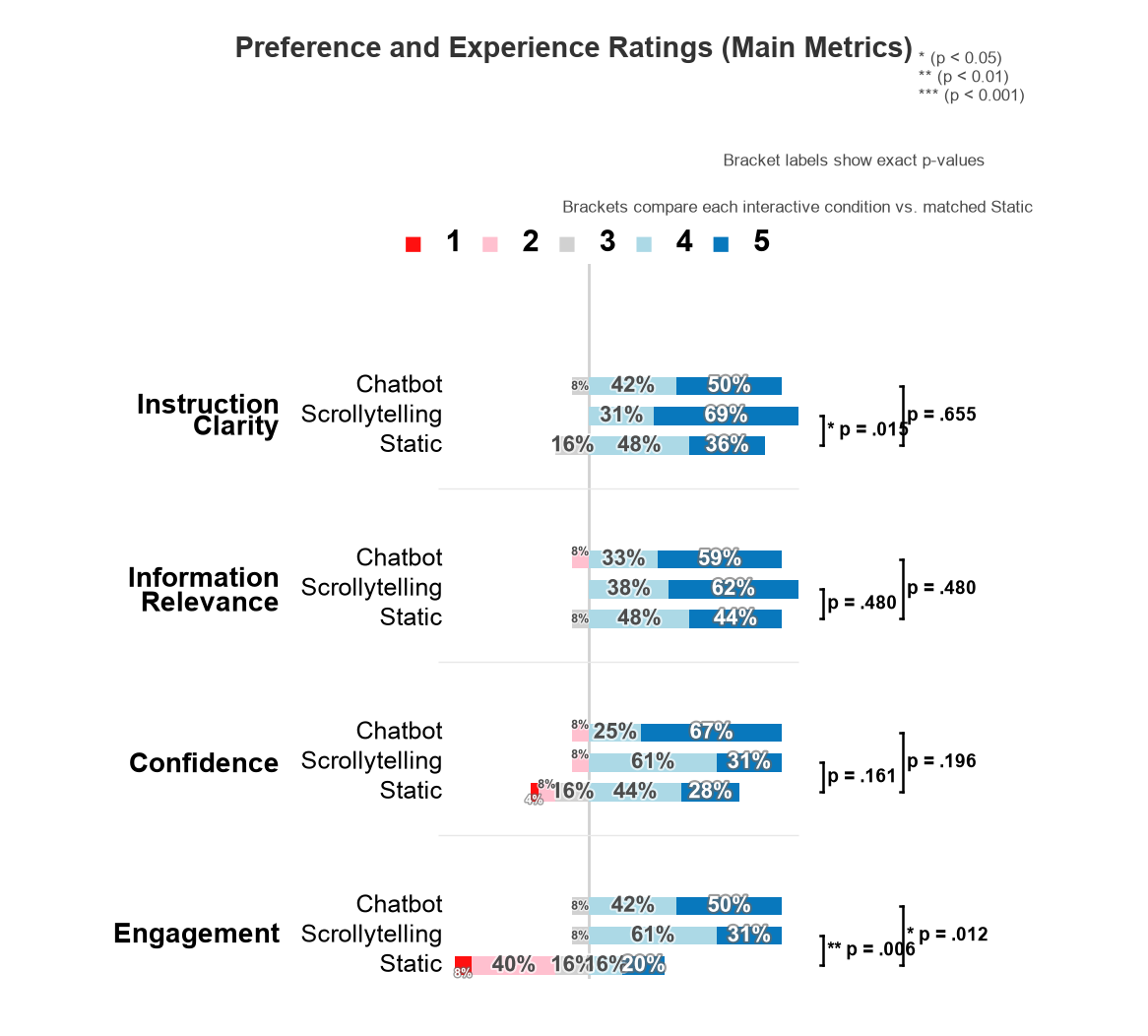} 
    \caption{A diverging bar chart showing participants' ratings of the three onboarding techniques across four core metrics: Engagement, Confidence, Information Relevance, and Instruction Clarity. Responses were recorded on a 5-point Likert scale, with metric-specific anchors listed in Table \ref{tab:Likert}. Percent labels are rounded within each row to sum to 100\%. Static bars aggregate all static responses; bracket labels report Wilcoxon Signed-Rank tests comparing each interactive condition with its participant-matched static subset. Chatbot versus scrollytelling comparisons were tested separately with the Mann-Whitney $U$ test and are reported in the Results section. Consistent with the main results, both interactive methods had significantly higher engagement ratings than the static condition, instruction clarity ratings were higher for scrollytelling and in the pooled interactive-versus-static analysis, and confidence ratings did not reliably differ across conditions.}
    \label{fig:likertResults}
\end{figure}

Confidence ratings were numerically higher in the overall interactive condition (\(M{=}4.32,\,SD{=}0.85\)) than in the static condition (\(M{=}3.84,\,SD{=}1.07\)), but this difference did not reach the conventional significance threshold ($p = .054$). When separated by technique, neither scrollytelling nor chatbot significantly differed from its matched static condition (Appendix Figure \ref{fig:likertResults}).

Instruction clarity ratings were also significantly higher in the overall interactive condition compared to the static condition ($p = .048$). When separated by technique, scrollytelling showed higher instruction clarity ratings than its matched static condition ($p = .015$), whereas chatbot did not ($p = .655$). Information relevance ratings did not significantly differ across conditions (all $ps > .05$).

In summary, when compared to static onboarding, \textbf{the pooled interactive condition was associated with higher guideline adherence, and both interactive techniques received higher engagement ratings.} Confidence ratings were numerically higher for the pooled interactive condition but did not reach the conventional significance threshold. Satisfaction and instruction clarity ratings were higher for scrollytelling relative to its matched static condition, while quiz scores and information relevance ratings did not vary significantly among onboarding techniques.

\subsubsection{Comparison of Scrollytelling and Chatbot Conditions}
When comparing the scrollytelling and chatbot interactive conditions, no significant differences were found across quiz scores or guideline adherence. Quiz scores were similar between scrollytelling (\(M{=} 4.76,\,SD{=} 0.43\)) and chatbot (\(M{=}4.75,\,SD{=}  0.45\)) conditions, and guideline adherence scores were comparable between the two interactive techniques as well.
For the onboarding experience ratings, no statistically significant differences were observed between chatbot and scrollytelling across confidence, engagement, information relevance, instruction clarity, or satisfaction ratings (all $ps > .05$; confidence: $p = .130$, engagement: $p = .428$, information relevance: $p = .776$, instruction clarity: $p = .296$, satisfaction: $p = .734$).

In summary, while both interactive techniques yielded similar outcomes in comprehension and creation tasks, no significant differences were found in onboarding experience ratings between chatbot and scrollytelling.


\subsection{Qualitative Results}

After completing the onboarding tutorials and creating visualizations, participants shared their preferences and perceptions of each tutorial. Their feedback indicated that while most participants preferred interactive onboarding, static onboarding was beneficial in some cases.

\subsubsection{Interactive Onboarding is Preferred for its Engagement, Bite-size Information, and Reported Ease of Recall}

After completing both static and interactive onboarding techniques, the majority of the participants ($n=20$, 80\%) preferred the interactive technique. When asked about the reason for this preference, they stated that the interactive onboarding kept them engaged and motivated to complete the whole content through user interaction, provided bite-size information, and made the information feel easier to retain.
First, participants described interactive onboarding techniques as more engaging ($n=7$). As one participant explained, \textit{"I feel like the information was similar, but scrolling was a little more engaging. So it was like, yeah, it took me shorter to remember"} (P20).
Participants said that the chatbot's approach of letting users select the next topic helped them stay focused and attentive. P12 pointed out that \textit{"it keeps your focus over there because you have to [decide what to click on], if you want to move next."}

The next advantage of interactive onboarding was that it structured the content as bite-sized steps of information ($n=6$), which was often cited as the reason for feeling confident about designing a visualization after the tutorial. Participants mentioned \textit{"it was very easy to understand in the chatbot because it was all like, broken down into very simple steps, helped me to understand both the data visualizations pretty easily"} (P2) and helped them focus on \textit{"limited scope"} (P6).
 As P1 explained, \textit{"In the scrollytelling tutorial, where it gave a step by step breakdown of what is supposed to be done, and because it was not all the information at once, it was easier to get through it."} This breakdown of content could have contributed to higher adherence to the design guidelines. As an illustration, P4 verbalized the guidelines step-by-step as they created a heatmap after going through the chatbot heatmap tutorial:\textit{``first off, based on the tutorial, my instant first step I feel like doing is creating the x and y axis"}; \textit{``Now, values are obviously too big, [...] So I'll probably put the scale to 100,000"}; \textit{``The minimum is this [17...] if we choose a color, we'll put the lightest colors at this point"}; etc.

On the other hand, participants felt the static onboarding format was difficult to complete due to visual and information overload, even though the same amount of information was presented by each onboarding technique. This was a common concern shared among several participants who stated, \textit{"I think the static one, my immediate reaction to text is like, oh my God, so much text"} (P17), \textit{"But when you read a word document, it feels like you're doing a huge task, and then it slowly starts to get, you know, your attention level just slowly dies down"} (P9), and \textit{"I think I was forced to sit and sift through all of the information in the static version versus the chatbot"} (P1).

Another advantage of interactive onboarding reported by participants was easier recall ($n=17$): \textit{"I was able to recall the instructions from both [static and chatbot onboarding techniques], but I felt like it was easier recalling instructions from the chatbot again, because it felt a little more engaging when I clicked certain instructions"} (P10). On the other hand, after the static onboarding, participants felt they were not able to remember some of the steps while creating the visualizations. This pattern was consistent with creation-task errors we observed in the static condition, such as missing legends next to heatmaps despite guideline recommendations.
After rating confidence level after the static tutorial as ``not confident at all," P1 elaborated, \textit{"textual was very text heavy, so I couldn't remember the steps, even like the steps to get to designing the heatmap and assigning the colors. It was like I was very confused at the end. [...]
I couldn't remember [...] how to assign colors in the legend and all of that. But in scrollytelling, I at least knew like the basics of what to do."}

The content layout in static onboarding, where images are interspersed between text, also made it more difficult for participants to map textual guidance to what was being illustrated in the visualization. P25 highlighted this friction, noting that when scrolling through a standard text document and \textit{"there is an image above and there is some text and then there is an image below, then sometimes you tend to get confused."} This disconnect might have hindered participants' ability to apply the design guidelines, and thus contributed to lower guideline-adherence scores. In contrast, the increased prominence of visual examples within interactive onboarding appeared to better support the application of guidelines. For example, P12 shared relying almost exclusively on visual memory during both creation tasks: \textit{"[I was] trying to think of what I saw and how it looked like instead of, not like... I don't think I recollected any of the text that was already there. Yeah, I referenced more to the images."}

\subsubsection{Benefits of Static Text-Based Onboarding}
Although most participants preferred interactive onboarding, they also reported the advantages of static text-based onboarding including the linear flow and the familiarity of the format.
Participants appreciated the simplicity of the static layout, noting the straightforward progression, which was \textit{``just basically reading, looking at the images, and understanding"} (P22). This familiarity of this linear structure allowed them to focus entirely on the content rather than the delivery method.
P3 found the novelty of the chatbot a distraction: \textit{"I was not used to chatbot learning [...] that's why I wasn't able to grasp all the information I needed for the task later on. But whereas in textual, I was very used to reading stuff. [...] I guess in chatbot, I felt that it was a lot. [Textual] was a simple flow"} (P3). Furthermore, the ability to view all instructions simultaneously was seen as a benefit as P23 noted, \textit{"I think the text is better as a new learner, where I can have the whole information every step in, like, one go."} Along with these advantages, participants mentioned it would be helpful if the text is written in a concise and clear way, and if a chatbot would answer their questions when needed alongside the static information.

\subsubsection{Two-way Interaction Offers More Flexibility, While One-way Interaction Is Reported as Less Effortful}
As the study design used a between-subject approach for the two interactive onboarding techniques, we could not ask for a direct comparison of the two techniques. However, our qualitative analysis of participants' feedback on each technique allowed us to extract the main similarities and differences.
Although the two interactive techniques did not differ significantly on measured onboarding experience ratings, participants distinguished them based on navigational flexibility, reported physical and mental effort, and the positioning of visualizations in relation to the text.

The chatbot (i.e., two-way interactive technique) had a feature to choose the next step the user wanted to learn, and participants liked having this navigation flexibility in their interaction. As noted by participants, \textit{"I think the chatbot would be more helpful, because you get more freedom over there. Like, you get to interact with it, you get to engage with the material that is over there"} (P21). Some participants also expressed a desire for an onboarding chatbot that could answer ad hoc questions (P11), even though our study prototype only supported preset prompts (i.e., no free-form queries).
The chatbot's ability to let users skip sections was appreciated, with one participant mentioning, \textit{"I feel that somehow the chatbot felt more interactive because you're selecting what you want to learn, and if you know everything, you can just skip it"} (P10).
Another advantage of the chatbot was that it required less physical effort, as expressed by one participant: \textit{"I'm not really a fan of scrolling too much. [...] if there is a thing that I can choose from, I just would prefer that so I don't have to like, scroll all the way and find what I want to learn"} (P7). However, some participants reported higher mental effort because they had to choose from a list of multiple options after each step, compared to scrollytelling (i.e., one-way interactive technique), where they simply followed a preset flow of information.

Some participants also highlighted scrollytelling onboarding's information presentation layout, where the visualization was placed on the right side of the text content as shown in Figure \ref{fig:scrollytelling}. As described by a participant, \textit{"I like the scrolling one better because I had the visualization right beside and those scrolling interactions for it much better and smoother than the document I guess"} (P17). In contrast, for chatbot and static onboarding, the visualization moved with the text, leading to instances where the visualization was cut off and participants had to scroll back to view it.
Participants shared that another benefit of the side-by-side information layout was that it allowed for the ``animation" of the visualization, where the visualization content could be updated in place ($n=4$). A participant noted, \textit{"I think the animation caught my eye more when it started happening I might quickly [go] to that and it was something that I noticed"} (P14).

\section{Discussion}
In this section, we answer our three research questions. First, we examine how static, scrollytelling, and chatbot onboarding compare in enhancing UX designers' ability to understand (RQ1) and design (RQ2) complex data visualizations.  We then examine how the interactive techniques (scrollytelling and chatbot) influence the overall onboarding experience for UX designers (RQ3). We also discuss the different dimensions of onboarding, design implications, and potential future research in this field.

\subsection{Effectiveness of Interactive Onboarding Techniques as Compared to Static Onboarding}
Both quantitative and qualitative results highlight the distinct benefits of interactive techniques like scrollytelling and chatbots over traditional static techniques. Prior research has found that interactive onboarding can support comprehension compared to static onboarding, although the difference was not statistically significant \cite{stoiber2022comparative,kwon2016comparative}. Similar to these earlier findings, our study did not find any significant difference in comprehension between interactive and static techniques although the average score in interactive conditions was slightly higher than that in static condition (RQ1). This may be due to a ceiling effect since all participants scored highly on the comprehension quiz regardless of the onboarding technique \cite{Garin2014}. More notably, the pooled interactive condition showed significantly higher guideline-adherence scores than static onboarding, addressing RQ2. This is a novel finding, as prior studies have not explored the effectiveness of visualization onboarding for visualization \textit{creation}, which is a useful skill to teach UX designers as they create dashboards. When comparing the two interactive techniques (RQ2, RQ3), scrollytelling and chatbot showed similar guideline-adherence scores and experience ratings.


Interview data capturing participants’ preferences and reflections on each onboarding technique further suggests that interactive onboarding provides broader reported benefits, including participants' reports of greater engagement and clearer instructions (RQ3), a numeric trend toward higher confidence ratings, and higher satisfaction for scrollytelling. These benefits may make interactive techniques useful for guiding users through complex tasks like visualization creation, where multiple guidelines must be learned and applied.
While the pooled interactive condition was associated with higher guideline adherence and both interactive methods received higher engagement ratings than static onboarding, confidence ratings were only numerically higher in the pooled interactive condition. No significant differences were observed between chatbot and scrollytelling. Some qualitative comments suggest that the active, dialogue-driven nature of a chatbot might have given users a greater sense of agency as they chose the aspects to learn about.

Based on the findings, we also note the importance of using an interactive onboarding format that requires attention and \textit{active} interaction to proceed, rather than one that allows for passive consumption of information. For illustration, we compare the interactive methods in our study to video-based tutorials, which is a popular onboarding strategy in educational systems \cite{lafreniere2013video,grossman2010video,pongnumkul2011video}. While video tutorials are non-static, video-based onboarding may not always actively engage learners as they typically allow for passive information absorption and thus restrict knowledge retention and application \cite{smith2022engagement}. Our participants described the interactive formats as more engaging and highlighted user-controlled pacing as a strength, which suggests that designers could augment passive, linear video tutorials with interactive components or support ``choose your own adventure" style branching when videos are used for onboarding.

\subsection{Design Dimensions of Visualization Onboarding}
Based on the findings, we discuss three key dimensions of visualization onboarding techniques that can support designers in developing effective onboarding materials.
\subsubsection{Structuring the Onboarding Narrative}
The interview results showed that the structure of onboarding information is critical for user engagement and learning. We build on prior research on narrative visualization that explored various formats of storytelling to engage viewers \cite{HeerVan,Segel2010Narrative,Hullman2013Narrative,Hullman2011Rhetoric}. Hullman et al. \cite{Hullman2013Narrative} presented and listed the advantages of six ways of storytelling: linear sequencing, parallelism, guided emphasis, reader-driven storytelling, grouping and structuring, animation, and transition techniques.
The three onboarding techniques in our study used various combinations of these narrative elements. Scrollytelling primarily supports a linear narrative with dynamic transitions and guided emphasis. Chatbots provide an interactive, reader-driven experience, allowing users to explore content non-linearly while still benefiting from guided emphasis. Static onboarding provides organized, non-interactive material in a familiar style, supporting basic linear learning.

Linear approaches can ensure foundational concepts are not overlooked through ``forced exposure." This could be beneficial for beginners, especially for learners at the stage of ``unconscious incompetence" \cite{Lynch2020AI}. At this stage, learners are unaware of their knowledge gaps and lack the domain vocabulary that is needed for productive exploration. In contrast, reader-driven approaches allow users to selectively access relevant information, which could improve efficiency and support reported confidence, as reflected in participants’ qualitative feedback about the two-way interactions in chatbot onboarding.


\textbf{Design Implication 1. Align narrative structure with user expertise or task context:}
These findings suggest that the narrative structure of visualization onboarding should be tailored to the user's knowledge level and situational constraints. Linear, step-by-step narratives are more appropriate for novice users to ensure foundational coverage, while reader-driven exploration better supports advanced users who need efficient, targeted access to information. Chatbot onboarding may also be more appropriate in time-sensitive contexts, such as preparing a visualization for an imminent stakeholder review or learning a tool under deadline pressure, as users can access the necessary information without traversing the full tutorial.
A hybrid design can combine these strengths by starting with a structured linear tutorial and gradually introducing interactive elements, such as chatbot-based support.

While our study used chatbots with predefined prompts, future work could explore Large Language Model (LLM)-based chatbots to facilitate a co-authored narrative. Prior work has shown that such systems help users understand complex visualizations by offering explanations, corrections, and personalized support through interactive discussions and visual techniques \cite{LLMVis}. Extending this to onboarding, LLM-driven chatbots could also assist in creating visualizations through helping with data interpretation, suggesting appropriate chart types, and checking designs against accessibility and best-practice guidelines.

\subsubsection{Designing Visual Content Layout and Cues}
Qualitative results showed that participants appreciated how scrollytelling displayed the visualization at a fixed location, allowing them to observe changes over time without reprocessing a new image at each step. Participants reported that this layout felt less overwhelming and required less mental effort to track compared to chatbot and static tutorials. Participants also expressed a desire for more visual cues that directly coordinate with the text to aid comprehension. This aligns with prior research, which showed that visual cues in online tutorials enhance learning by directing attention, improving retention, and help manage complexity, especially when combined with text \cite{Wang2020Impacts}. Additionally, they can make learning more efficient and engaging, particularly for beginners and those with limited computer literacy \cite{Jamet2016Enhancing,Wang2020Impacts,Lin2016Effects,Moon2020The,Aftab2020Remo}. Visual cues can be internal cues that modify original data characteristics (e.g., brightness, color, transparency) or external cues that add extra components (e.g., outlines, annotations, arrows) \cite{kong2017internal}. Internal cues reduce clutter but may obscure detail, whereas external cues provide clarity but risk visual overload. Designers must balance these approaches to support clarity without distraction.

\textbf{Design Implication 2. Keep visual references in a fixed location and use visual cues: }
Visualization onboarding should maintain a stable visual reference point to reduce reported visual-search effort and shifts in attention.
When possible, visualizations should remain in a fixed position in the layout while textual explanations update, enabling users to incrementally build understanding. While this is inherent in scrollytelling, we can achieve a similar effect in chatbot onboarding by keeping the visualization in a persistent side panel rather than embedding it within the chat stream.
Designers can further help users map abstract text-based design guidelines onto concrete visual changes by using visual cues to highlight the relevant region of a chart, animating transitions between states, or progressively revealing components. For example, in a pie chart tutorial on monthly expenses, we can display the chart alongside the textual instruction and use arrows (external visual cue) or modify the brightness (internal visual cue) to point out segments like "Rent" or "Groceries" as they are discussed, making the connection between text and visualization clear.

Future research could explore additional forms of interactivity that link text to its visual counterpart, such as data visualization brushing, where hovering over a highlighted section of the text updates the content shown on the visualization panel and vice versa.
Future research could examine how different cueing strategies impact retention, engagement, and adaptive learning in visualization onboarding. Prior research also showed that pairing visual cues with audio narration can further guide attention while reducing visual complexity \cite{HKongVisual}, offering another promising direction of multimodal guidance for visualization onboarding.

\subsubsection{Balancing Interactivity and Reported Mental Load}
Stroiber et al. investigated how educational theories on cognitive load can inform visualization onboarding design to enhance user knowledge and engagement \cite{stoiber2023design}. \textit{Cognitive Load Theory} explains how the limited capacity of working memory constrains learning \cite{sweller2011cognitive}. It distinguishes between intrinsic load, arising from the inherent complexity of the material, and extrinsic load, caused by how information is presented and often unnecessarily increased by poor design.
Although our study did not objectively measure cognitive load, our qualitative findings are consistent with this theoretical framework, suggesting that the presentation style of onboarding techniques might influence reported mental effort.
Participants' accounts suggest that static onboarding resulted in higher reported mental effort since participants consistently shared that its dense design felt visually demanding, as opposed to interactive formats that presented information in segmented steps.
The option selection in chatbot onboarding introduced a more nuanced trade-off. While allowing users to skip unnecessary steps increased efficiency, it also increased reported mental effort by requiring users to decide their learning path. This effect was amplified when option labels contained unfamiliar or vague keywords.
This suggests that although interactive interfaces can support reported engagement, they can also increase the reported effort associated with decision-making.

\textbf{Design Implication 3. Manage Navigational Flexibility to Reduce Reported Effort: }
Designers should consider how offering navigational flexibility might inadvertently increase reported mental effort. To mitigate this potential increase in chatbot onboarding, designers could either provide a brief explanation of the options at the beginning of the tutorial or make the option bubble labels more descriptive (e.g., changing "x-axis" to "x-axis (bottom horizontal line)", "Data" to "Understand Data Series"). However, each approach has a limitation, where the user can forget the explanation if presented only at the beginning, while in-bubble explanations increase visual complexity.
Simplifying the decision-making by removing already visited options may be an effective alternative, since Hick's Law states that the decision time is directly proportional to the number of options \cite{hick1952rate}. Advanced approaches, such as LLM-based chatbots, can further adapt guidance by dynamically adjusting options and explanations based on user interactions. For instance, the chatbot could detect when a user hesitates or frequently revisits an option and proactively filter irrelevant options or offer clarifications and suggestions.

The findings also suggest that well-structured, highly guided experiences with limited navigational flexibility (e.g., scrollytelling or concise static references) may be more suitable for high-stress circumstances and novice users, where increasing mental effort may have more serious consequences.
However, as our study did not directly measure cognitive load, future work is needed to validate these effects.

\section{Limitations and Future Work}
Our study featured a relatively small demographically skewed sample (18 students, 7 UX professionals; all with at least a college-level education), which may limit the generalizability of our findings to more diverse or expert populations. Future research should employ a more balanced and heterogeneous sample to determine if professional experience significantly moderates the effectiveness of interactive onboarding.
Next, visualization literacy was measured using a self-reported single-item scale (e.g., \textit{``How would you rate your data visualization knowledge on a scale of one to five?''}) rather than a validated instrument (e.g., VLAT), which may not accurately reflect actual visualization literacy level. Future studies are needed to confirm that the findings hold when objective visualization literacy is measured.
Furthermore, the lack of statistically significant differences between chatbot and scrollytelling across onboarding experience ratings may stem from both limited statistical power and the constrained technical nature of the chatbot itself.

As mentioned in the system description, the chatbot in this study was a rule-based dialog system rather than an AI-based agent. As a result, its effectiveness may have been limited compared to more adaptive AI-based systems. Future work can explore the effectiveness of various interactive elements that can be incorporated into an onboarding tutorial including LLM-based chats, GIFs, short videos, and gamification. Finally, our study was limited to heatmaps and treemaps and focused on immediate performance. Longitudinal studies are needed to determine the long-term retention and generalizability of these techniques across more complex visualization types.

\section{Conclusion}
Our study compared three onboarding techniques --- static, chatbot, and scrollytelling --- and found that the pooled interactive condition was associated with higher guideline-adherence scores during visualization creation, and that both interactive techniques received higher engagement ratings than the static technique. Instruction clarity ratings were significantly higher when the two interactive conditions were pooled. Although no significant differences were found between chatbot and scrollytelling in performance or onboarding experience ratings, the two interactive techniques offered distinct interaction affordances. Based on these findings, we recommend employing a structured narrative with step-by-step explanations, properly aligning text and images, and choosing or combining static and interactive techniques based on user proficiency and task context.

\bibliographystyle{abbrv}
\bibliography{finalb}

\newpage

\section{Appendix}

\begin{table}[!htbp]
  \centering
  \caption{Likert scale questions used in the survey and the corresponding response anchors.}
  \label{tab:Likert}
  \small

  \setlength{\tabcolsep}{5pt}
  \renewcommand{\arraystretch}{1.25}

  \newcommand{\numbox}[1]{%
    \begingroup\setlength{\fboxsep}{1.2pt}%
    \fcolorbox{gray!40}{gray!12}{\strut\makebox[1.1em][c]{\textbf{#1}}}%
    \endgroup
  }

  \rowcolors{3}{gray!5}{white}

  \begin{tabular}{>{\raggedright\arraybackslash}p{0.38\linewidth}
                  >{\raggedright\arraybackslash}p{0.58\linewidth}}
    \toprule
    \rowcolor{gray!12}
    \textbf{Metric \& question} & \textbf{Scale anchors (1--5)} \\
    \midrule

    \textbf{Instruction clarity}\\\textit{How clear were the instructions provided in the onboarding tutorial?}
    & \numbox{1} Very confusing; \numbox{2} Confusing; \numbox{3} Neutral; \numbox{4} Clear; \numbox{5} Very clear \\

    \textbf{Information relevance}\\\textit{To what extent did the tutorial cover information relevant to your learning needs?}
    & \numbox{1} Not relevant at all; \numbox{2} Slightly relevant; \numbox{3} Neutral; \numbox{4} Relevant; \numbox{5} Very relevant \\

    \textbf{Engagement}\\\textit{How engaging was the onboarding tutorial?}
    & \numbox{1} Not engaging at all; \numbox{2} Slightly engaging; \numbox{3} Neutral; \numbox{4} Engaging; \numbox{5} Very engaging \\

    \textbf{Confidence}\\\textit{How confident do you feel about designing a visualization after completing the tutorial?}
    & \numbox{1} Not confident at all; \numbox{2} Slightly confident; \numbox{3} Neutral; \numbox{4} Confident; \numbox{5} Very confident \\

    \textbf{Satisfaction}\\\textit{Overall, how satisfied are you with the onboarding tutorial?}
    & \numbox{1} Extremely dissatisfied; \numbox{2} Somewhat dissatisfied; \\
    & \numbox{3} Neither satisfied nor dissatisfied; \numbox{4} Somewhat satisfied; \numbox{5} Extremely satisfied \\

    \bottomrule
  \end{tabular}
\end{table}

\vspace{3em}

\begin{figure}[!htbp]
    \centering
    \includegraphics[width=\linewidth]{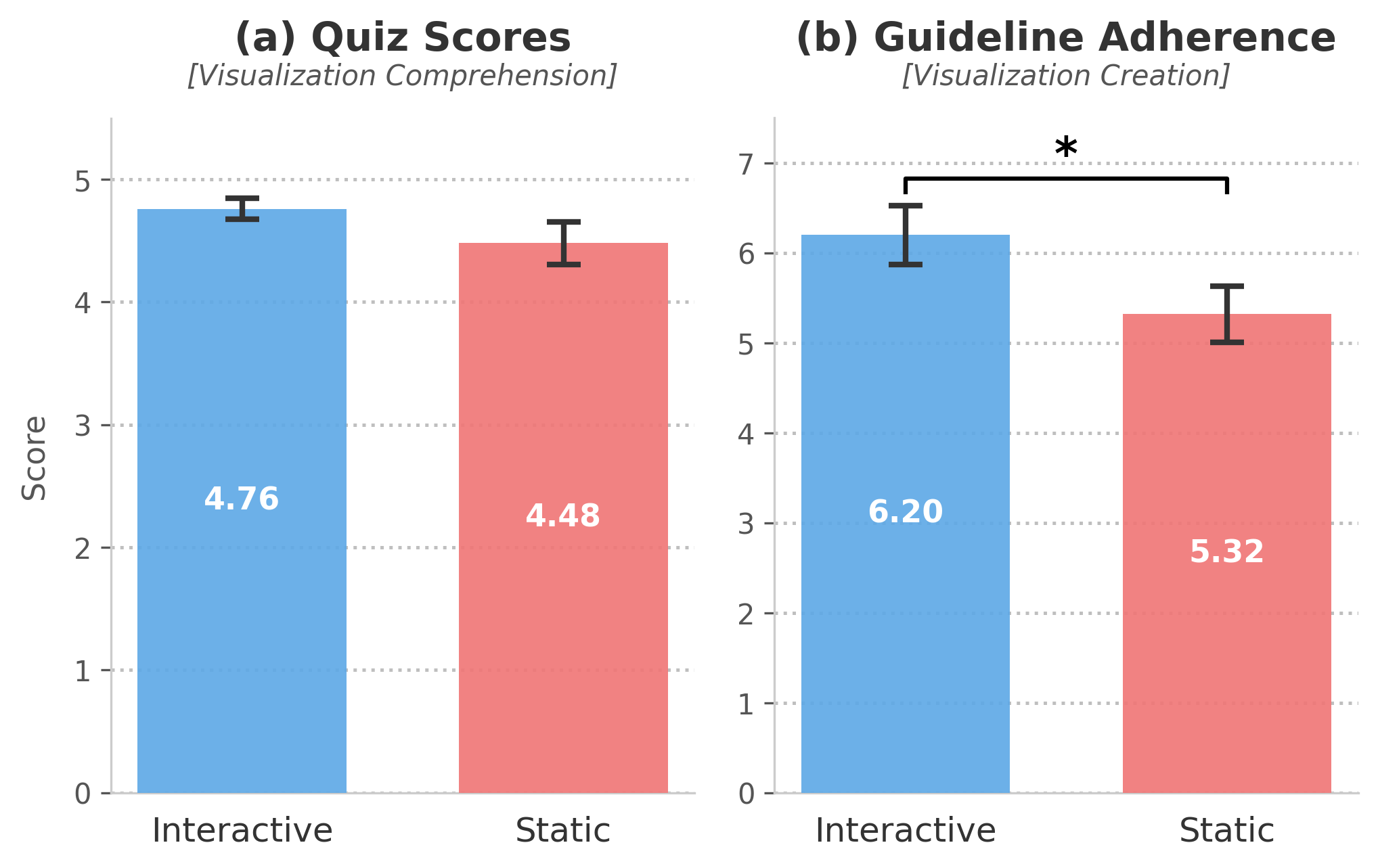}
    \caption{These bar charts show (a) the quiz scores (measured out of 5) and (b) guideline adherence scores (measured out of 8) for participants after following static versus pooled interactive onboarding. Error bars represent the Standard Error of the Mean (SEM). The asterisk (*) indicates a statistically significant difference ($p < .05$), with pooled interactive onboarding associated with higher guideline-adherence scores as shown in (b).}
    \label{fig:quantResults}
\end{figure}






\begin{figure}[!htbp]
    \centering
    \includegraphics[width=\linewidth]{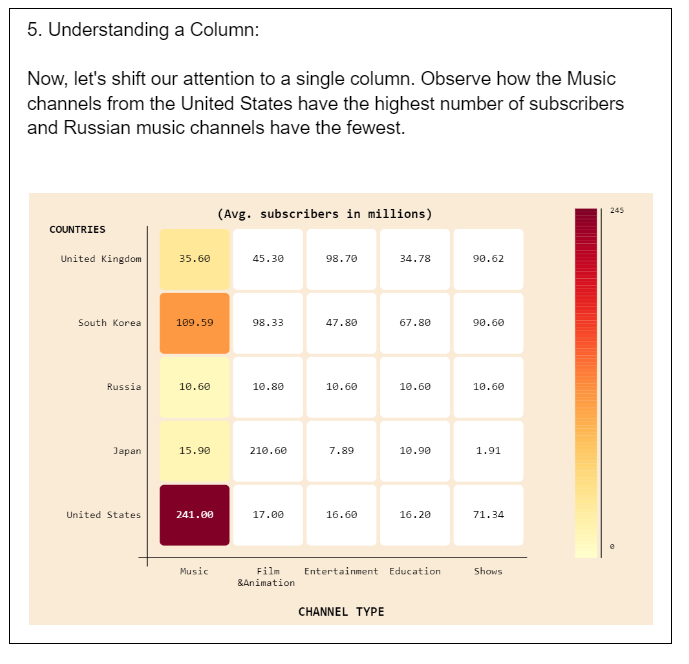}
        \caption{A step from the static tutorial featuring explanatory text at the top with an image related to that text directly below it.}
    \label{fig:static_left}
\end{figure}

\begin{figure}[!htbp]
    \centering
    \includegraphics[width=\linewidth]{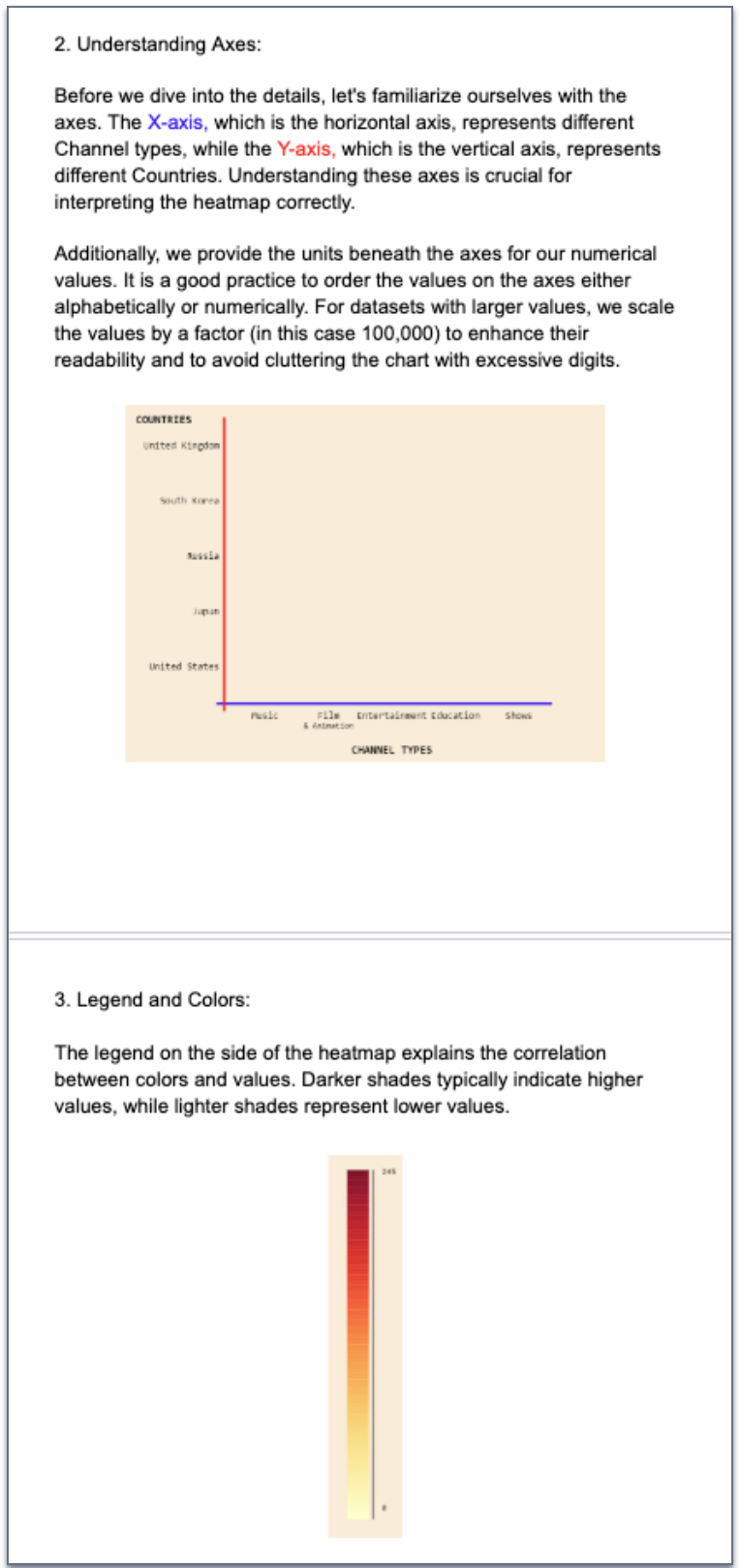}
    \caption{The static tutorial follows a linear flow, with sequential steps displayed one after the other, each containing text and an image. Full versions of the static tutorials for both treemaps and heatmaps are provided in the supplementary materials (PDF).}        \label{fig:static_right}
\end{figure}

\begin{figure}[!ht]
    \centering
    \begin{subfigure}{0.4\textwidth}
    \centering
    \includegraphics[width=\textwidth]{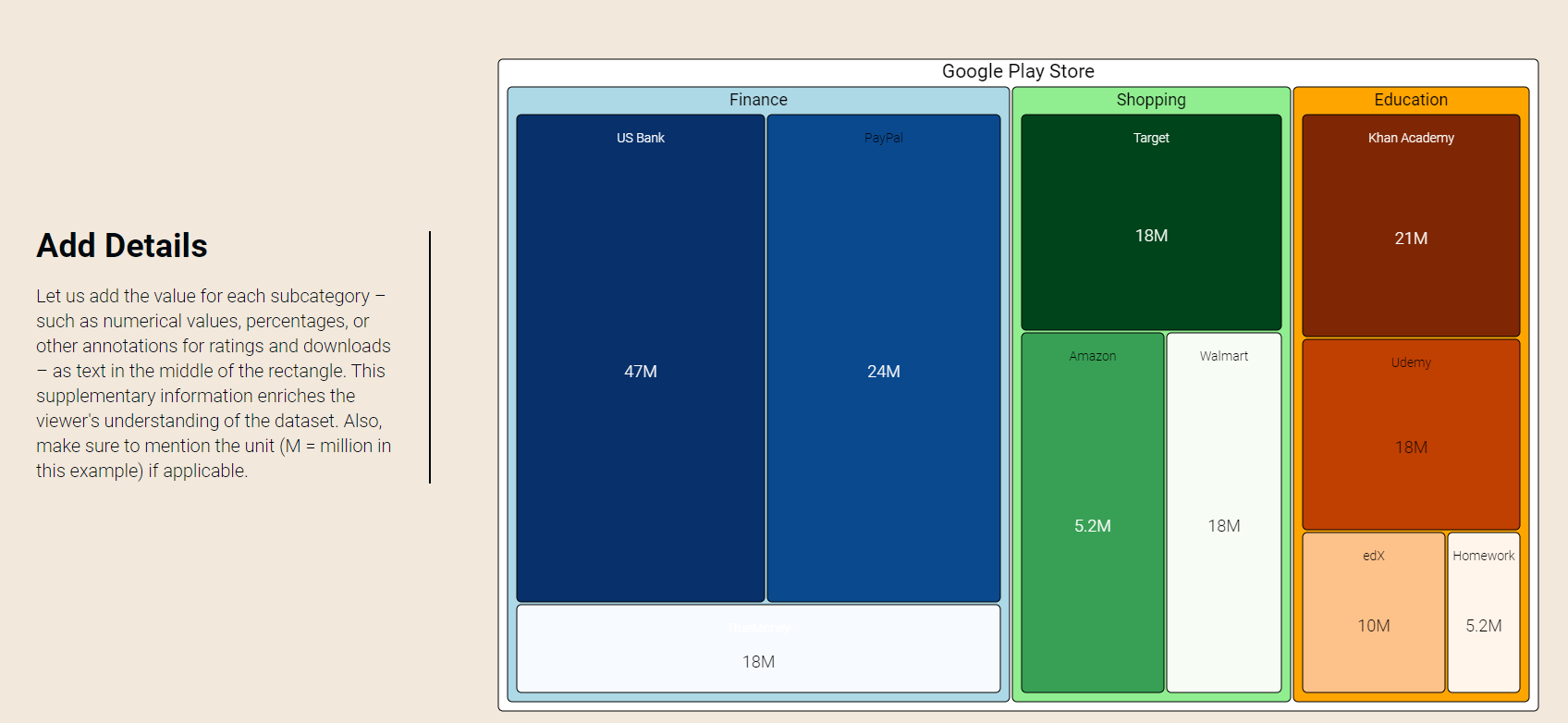}
    \caption{Scrollytelling tutorial with text explaining a step on the left and the animation on the right.}
    \label{fig:Stop}
\end{subfigure}
\vspace{1em} 
    \begin{subfigure}{0.4\textwidth}
        \centering
        \includegraphics[width=\textwidth]{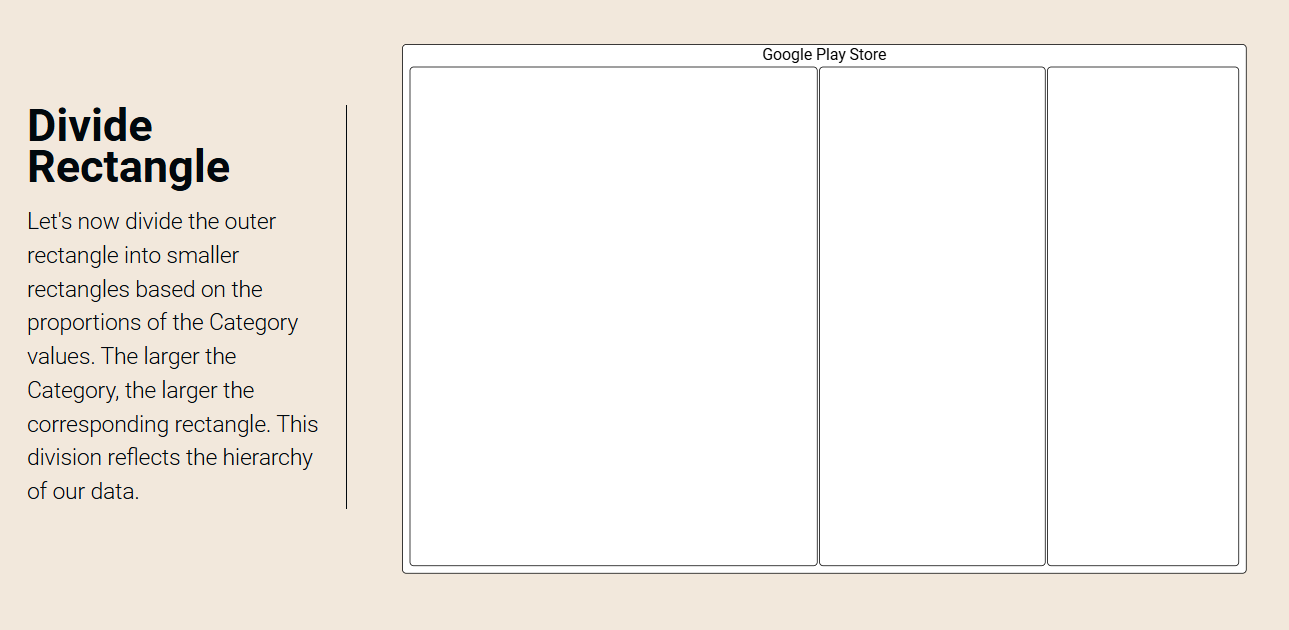}
        \caption{The user initially views the outer rectangle divided into categories before scrolling.}
        \label{fig:Sbottom_left}
    \end{subfigure}
    \hfill
     \begin{subfigure}{0.4\textwidth}
        \centering
        \includegraphics[width=\textwidth]{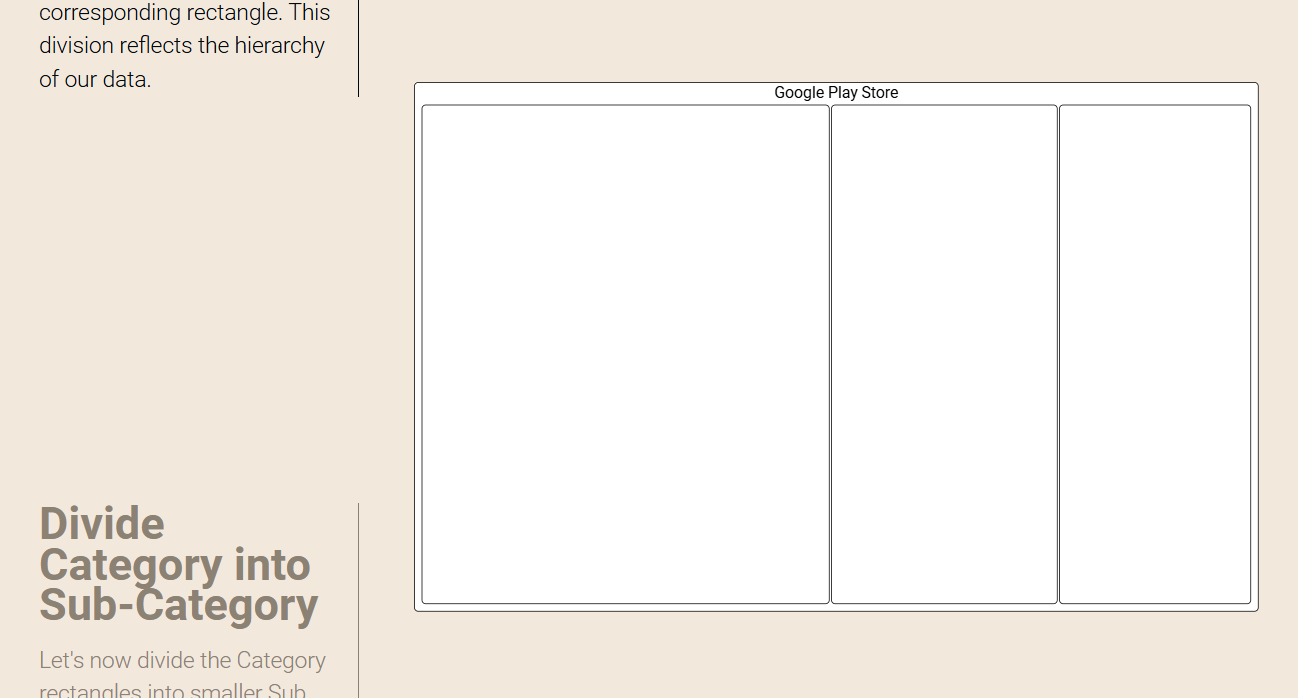}
        \caption{As the user scrolls, the text and images change, creating an animated effect.}
        \label{fig:Sbottom_mid}
    \end{subfigure}
    \hfill
    \begin{subfigure}{0.4\textwidth}
        \centering
        \includegraphics[width=\textwidth]{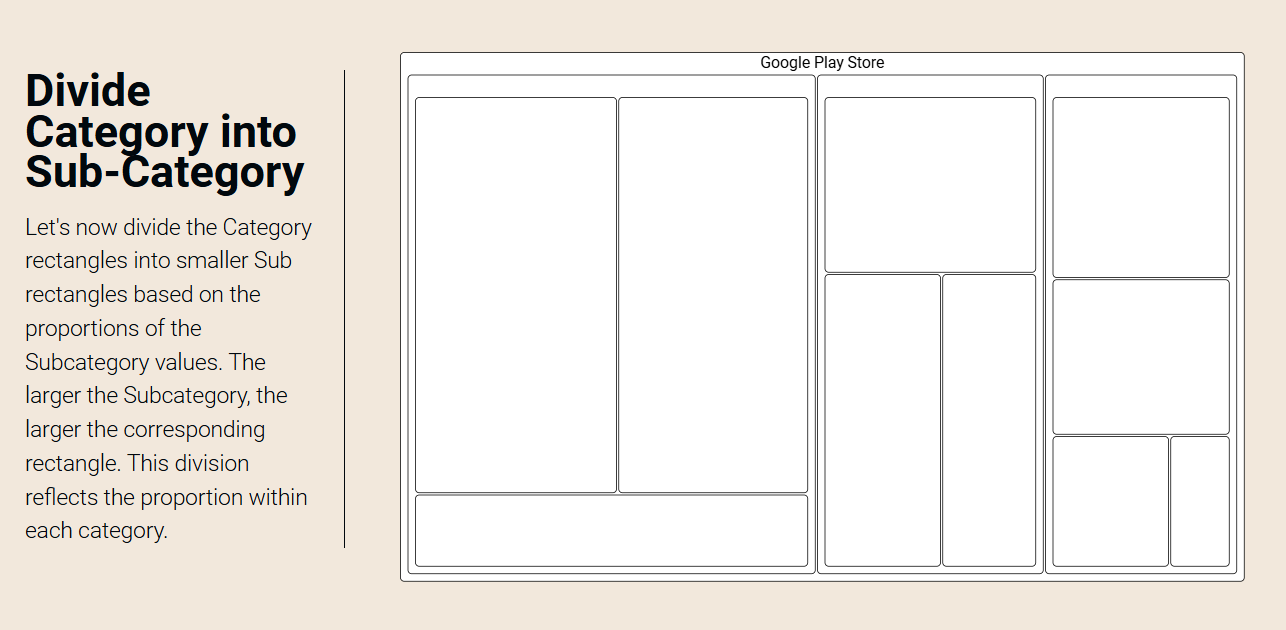}
        \caption{Further scrolling reveals the categories being subdivided into subcategories.}
        \label{fig:Sbottom_right}
    \end{subfigure}

    \caption{Fig (a) showing the final step with the completed visualization, Fig (b) - Fig (d) showing the dynamic transition between the steps in the scrollytelling tutorial.}
    \label{fig:scrollytelling}
\end{figure}

\end{document}